\documentclass[a4paper]{article}

\usepackage{INTERSPEECH2022} 
\usepackage{amsmath,amsmath }
\usepackage{multirow,cite}
\usepackage{graphicx}

\title{Scoring of Large-Margin Embeddings for Speaker Verification: Cosine or PLDA?}
\name{Qiongqiong Wang, Kong Aik Lee, Tianchi Liu \thanks{Tianchi Liu is also with Department of Electrical and Computer Engineering, National University of Singapore, Singapore.}}

\address{
  Institute for Infocomm Research (I$^2$R), A$^\star$STAR, Singapore}
\email{\{wang\_qiongqiong;lee\_kong\_aik;liu\_tianchi\}@i2r.a-star.edu.sg}

\begin{document}

\maketitle
\begin{abstract}
\vspace*{-2mm}
The emergence of \emph{large-margin softmax cross-entropy losses} in training deep speaker embedding neural networks has triggered a gradual shift from parametric back-ends to a simpler cosine similarity measure for speaker verification. Popular parametric back-ends include the \emph{probabilistic linear discriminant analysis} (PLDA) and its variants. This paper investigates the properties of margin-based cross-entropy losses leading to such a shift, and aims to find scoring back-ends best suited for speaker verification. In addition, we revisit the pre-processing techniques which have been widely used in the past and assess their effectiveness on large-margin embeddings. 
Experiments on the state-of-the-art ECAPA-TDNN networks trained with various large-margin softmax cross-entropy losses
show a substantial increment in intra-speaker compactness making the conventional PLDA superfluous. In this regard, we found that constraining the within-speaker covariance matrix could improve the performance of the PLDA. It is  demonstrated through a series of experiments on the VoxCeleb-1 and SITW core-core test sets with  $40.8\%$ equal error rate (EER) reduction and $35.1\%$ minimum detection cost (minDCF) reduction. It also outperforms cosine scoring consistently with reductions in EER and minDCF by $10.9\%$ and $4.9\%$, respectively. 
\end{abstract}

\vspace*{-2mm}

\section{Introduction}
\vspace*{-1mm}
Automatic speaker verification (ASV) is the process to verify whether a given speech utterance is from a specific speaker or not. I-vector embedding \cite{dehak11} followed by \emph{probabilistic linear discriminant analysis} (PLDA) \cite{prince07,ioffe06} was dominant in ASV for a long time until recent years when ASV started to benefit from deep learning. The use of deep neural networks (DNNs) has been investigated to replace individual components along the ASV pipeline, including the front-end feature extraction  \cite{Snyder18,Tang19}, back-end modeling \cite{Chien17}, and the entire pipeline in an end-to-end manner \cite{Li17, Rohdin20}. Among these, using DNNs to extract discriminative speaker embeddings has been shown to be the most viable and effective. Therefore, recent works in ASV have focused on building network architectures that produce embedding vectors with improved speaker representations\cite{Snyder18,Desplanques20, lee21,Liu22}.

A  DNN for extracting an utterance-level speaker embedding consists of three modules:
(1) a frame-level feature encoder, (2) a pooling layer, and (3) utterance-level representations. The input to the first module is a sequence of acoustic features, e.g., \emph{mel-frequency cepstral coefficients} (MFCCs) and filter-bank coefficients. After considering relatively short-term acoustic features, this module outputs intermediate representations. Various neural network architectures have been used as the encoder, e.g., the \emph{time-delay neural network} (TDNN) \cite{Snyder18}, convolutional neural network (CNN) \cite{Nagrani17}, 
LSTM \cite{Bhattacharya17}, the incorporation of LSTM to TDNN \cite{Tang19} or gated recurrent unit (GRU) \cite{Li17}.
The goal of this module is to extract more comprehensive speaker information. The second module converts variable-length frame-level intermediate features into a single fixed-dimensional vector by a temporal pooling. In addition to the most basic average statistics pooling, attention mechanism \cite{Zhu18,Okabe18,zhu21c_interspeech} is commonly used to form  weighted statistics focusing  on essential frames and in turn become more speaker discriminative. The third module stacks several fully-connected layers including one bottleneck layer used to extract utterance-level speaker embeddings with a fixed dimension in the testing phase. During training, the output nodes correspond to the set of speaker IDs in the training data. A softmax function is commonly used to constrain the predicted outputs so that they sum to one, and
a cross-entropy (CE) loss is used to measure the network performance.

Good speaker embeddings should be discriminative between different speakers and compact within the same speaker. Embeddings learned using the conventional softmax CEloss, however, are optimized for only inter-speaker discrepancy.
To address this issue, margin penalties have been introduced to the so-called large-margin softmax CE loss \cite{fwang18,hwang18,deng18}, to simultaneously enhance the intra-class compactness and inter-class discrepancy. In this paper, we refer to the embeddings extracted from networks trained with margin penalties as the large-margin embeddings.

The emergence of large-margin embeddings has triggered a gradual shift from parametric back-ends, such as the PLDA, to a simpler cosine similarity measure \cite{Liu19,Zhou20}.  One possible reason is that a PLDA model decomposes the total variability into within and between-speaker covariance matrices \cite{prince07,ioffe06}. The intra-speaker compactness of the large-margin embeddings makes the within-speaker variability modeling no longer essential. However, as we noted, there is no prior experimental analysis. The goal of this paper is three-fold: 
(1) to study the properties of large-margin embeddings with respect to their predecessors, and to find
(2) suitable scoring back-ends and
(3) pre-processing techniques best suited for large-margin embeddings.

The paper is organized as follows.
Section 2 reviews the large-margin softmax CE loss, 
as well as cosine similarity and PLDA back-ends.
Section 3 introduces our investigations and motivations.
Section 4 shows the experimental setup and results.
Section 5 provides a summary of our work.
\vspace*{-1mm}
\section{Large-Margin Embeddings for ASV}
\vspace*{-1mm}
\subsection{Softmax and Large-Margin Softmax}
\vspace*{-1mm}
\subsubsection{Softmax Cross-Entropy Loss}
\vspace*{-1mm}
The softmax function is often used as an activation function 
to calculate the relative probabilities to target classes in multi-way classification tasks.
The cross-entropy (CE) loss could be calculated as:
\begin{equation}
    L_1 = -\frac{1}{N}
    \sum_{i=1}^{N} \log 
    \frac{ e^{\mathbf{W}_{y_i} ^T \mathbf{x}_i+b_{y_i}}}
    {\sum_{j=1}^C e^{{\mathbf{W}_j ^T \mathbf{x}_i+b_j}}      }
    \label{eq:sm}
\end{equation}
where 
$N$ is the batch size, 
$C$ is the number of speakers in the training set, 
$\mathbf{x}_i\in \mathbb{R}^d$ is the embedding representation of the $i$-th utterance,
belonging to $y_i$-th class. 
The vector $\mathbf{W}_j\in \mathbb{R}^d$ denotes $j$-th column of the weight matrix $\mathbf{W} \in \mathbb{R}^{d\times C}$ while 
$b_j \in \mathbb{R}^n$ is the corresponding bias term. 
The softmax function constrains the total probabilities to all the classes as 1,  
which helps training converge more quickly than it otherwise would.
The expression ${\mathbf{W}_{y_i} ^T \mathbf{x}_i}$ in the numerator of (\ref{eq:sm}) is equal to 
$  \|{ \mathbf{W}_{y_i} } \| \|\mathbf{x}_i\| \cos(\theta_{y_i}) $,
with the angle $\theta_{y_i}$ between the vectors  $\mathbf{W}_{y_i}$ and $\mathbf{x}_i$.
A modified softmax CE loss \cite{ranjan17,liu17} further normalizes the individual weight vector  $\|\mathbf{W}_{j}\| = 1$,
normalizes the embedding vector  $\|\mathbf{x}_i\|$ and re-scales to $s$, and discards the bias term:
\begin{equation}
    L_2 = -\frac{1}{N}
    \sum_{i=1}^{N} \log 
    \frac{e^{s \cdot \cos(\theta_{y_i})}}
    {\sum_{j=1}^C e^ {s\cdot \cos({\theta_j})}      }.
     \label{eq:sm1}
\end{equation}
The modification enables the network to directly optimize angles and learn angularly distributed features, but not necessarily more discriminative ones  \cite{liu17}.
  
\subsubsection{Large-Margin Softmax Cross Entropy Loss}
Since angles are used as the distance metric in (\ref{eq:sm1}), 
various techniques were introduced to incorporate margin penalties
in order to enhance the speaker-discriminative power.
They can be summarized with an angular function \cite{deng18}
\begin{equation}
\psi(\theta_{y_i}) = cos(m_1\theta_{y_i}+m_2) -m_3
\end{equation}
where $m_1$, $m_2$ and $m_3$ are the three margin penalties. 
Therefore, the larger margin softmax cross-entropy (CE) loss is 
\begin{equation}
    L_3 = -\frac{1}{N}
    \sum_{i=1}^{N} \log 
    \frac{e^{s \cdot \psi(\theta_{y_i})}}
    {e^{s \cdot\psi(\theta_{y_i})} +\sum_{j=1,j \neq i}^C e^{s \cdot\cos({\theta_j})}      }.
    \label{eq:largesm}
\end{equation}
The margins $\{m_1$, $m_2$, $m_3\}$ can be used simultaneously \cite{deng18} or individually \cite{liu17,fwang18,hwang18,deng18},
in which (\ref{eq:largesm}) is
denoted, respectively, as the angular softmax (A-Softmax) \cite{liu17}
\begin{equation}
\psi(\theta_{y_i}) = cos(m_1\theta_{y_i}),
\end{equation}
the additive angular margin softmax (AAM-Softmax or ArcFace) \cite{deng18}
\begin{equation}
\psi(\theta_{y_i}) = cos(\theta_{y_i}+m_2),
\end{equation}
and the additive margin softmax (AM-Softmax)\cite{fwang18}
\begin{equation}
\psi(\theta_{y_i}) = cos(\theta_{y_i}) -m_3.
\label{eq:arcsm}
\end{equation}
The margin penalties enforce intra-class compactness and inter-class discrepancy.
This corresponds to a reduced within-speaker variability and a larger between-speaker variability 
in speaker recognition terminology. 
We refer to this class of representation as large-margin embeddings in this paper.

\subsection{Speaker Verification}

Speaker verification can be accomplished by calculating the similarity between the two speaker embeddings 
corresponding to an enrollment and test speech. 
To this end, a simple cosine distance measurement can be used.
Alternatively, a more sophisticated scoring back-end can be trained such as the \emph{probabilistic linear discriminant analysis} (PLDA).


\subsubsection{Cosine Similarity}
Cosine similarity scoring is a computationally efficient method in many verification tasks. 
When it is applied to speaker verification, the cosine of the angle between the enrollment ($\phi_e$) and test ($\phi_t$) embeddings is used as the decision score
\begin{equation}
s(\phi_{e},\phi_{t}) = \frac{<\phi_{e},\phi_{t}>}{\|\phi_{e}\|\|\phi_{t}\|}.
\label{eq:cos}
\end{equation}
This technique has an advantage that no training is required. 
Scoring is performed directly in the speaker embedding space.

\subsubsection{PLDA}
\label{sssec:plda}
As opposed to cosine similarity measure, PLDA is a supervised method where speaker labels are necessary to train a PLDA model. There are multiple PLDA variants \cite{prince07, ioffe06, Garcia-Romero11, Brummer10}. Here we focus on the formulation reported in \cite{prince07}, which is widely used in speaker recognition \cite{kenny10, Lee19}.

Let $\mathbf{\phi}$ be an embedding vector which we assume follows a Gaussian distribution \cite{bishop06,prince07,ioffe06}:
%
\begin{equation}
p(\mathbf{\phi | h,x}) = \mathcal{N}(\bf{\phi} | \mathbf{\mu} + \bf{Fh}+ \bf{Gx}, \bf{\Sigma} ),
\label{eq:gplda}
\end{equation}
where $\mathbf{\mu}\in \mathbb{R}^\mathit{d}$ is the global mean.
The matrices $\bf{F}\in \mathbb{R}^{\mathit{D\times d}}$ and $\bf{G}\in \mathbb{R}^{\mathit{D\times D}}$ are, respectively, the speaker and channel loading matrices, and $\mathbf{\Sigma}$ models the residual variances and is constrained to be a diagonal matrix. The vectors $\bf{h}$ and $\bf{x}$ are the latent speaker and channel variables, respectively. Integrating out the latent variables, we arrive at the following marginal density
\begin{equation}
    p({\phi})=\mathcal{N}(\phi|\bf{\mu}, \bf{\Phi}_\mathrm{B} + \bf{\Phi}_\mathrm{w} )
    \label{eq:marginal_density}
\end{equation}
where $\{\mathbf{\Phi}_\mathrm{B}, \bf{\Phi}_\mathrm{W}\}$ are the between and within-speaker covariance matrices given by
\begin{equation}
\bf{\Phi}_\mathrm{B}  = \bf{FF}^ \mathrm {T},
\bf{\Phi}_\mathrm {W} = \bf{GG}^ \mathrm {T} +\bf { \Sigma}.
\vspace*{-1mm}
\end{equation}
%
In the testing phase, the log-likelihood score between the enrollment ($\phi_e$) and test ($\phi_t$) embeddings is calculated as 
\begin{equation}
s(\phi_{e},\phi_{t})=\log \frac{p(\phi_e,\phi_t)}{p(\phi_e)p(\phi_t)}.
\end{equation}
Here, the joint likelihood in the numerator can be computed as
\begin{equation}
\scriptstyle
p(\phi_e,\phi_t)=\mathcal{N}\left(\begin{bmatrix}
           \phi_e \\
           \phi_t
         \end{bmatrix} \middle\vert
         \begin{bmatrix}
           \mu \\
           \mu
         \end{bmatrix},
         \begin{bmatrix}
           \bf{\Phi}_\mathrm{B} + \bf{\Phi}_\mathrm{w} & \bf{\Phi}_\mathrm{B}\\
           \bf{\Phi}_\mathrm{B}& \bf{\Phi}_\mathrm{B} + \bf{\Phi}_\mathrm{w}
         \end{bmatrix}
         \right),
\end{equation}
while the likelihood $p(\phi_e)$ and $p(\phi_t)$ in the denominator are evaluated using (\ref{eq:marginal_density}). It is evident that PLDA scoring involves the explicit use of between and within covariance matrices, which is absent in cosine scoring.

\begin{figure}[t]
\hspace*{-0.7cm}
\includegraphics[width=1.17\columnwidth]{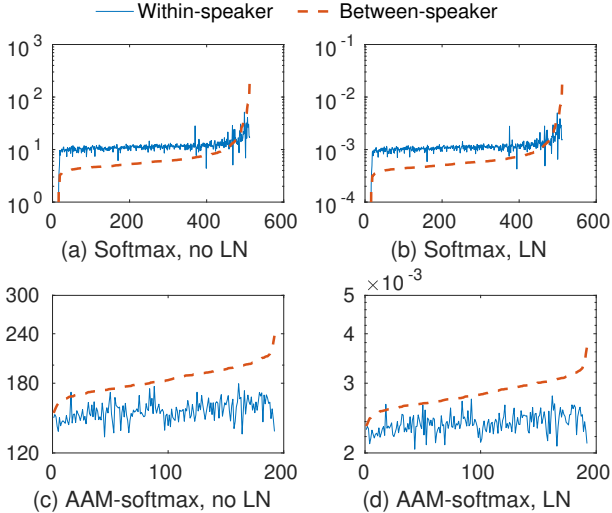}
\caption{\it Diagonal elements of the between and within-speaker covariance matrices of 
(a) conventional x-vector embeddings derived from a TDNN trained with softmax CE loss, 
(b) LN processed conventional x-vector embeddings,
(c) large-margin embeddings derived with an ECAPA-TDNN trained with AAM-Softmax CE loss, and 
(d) LN processed large-margin embeddings.
Values are sorted according to the between-speaker covariance matrices, and shown in log scale.}
\label{fig:cov}
\end{figure}
\begin{figure}[t]
\centering
\includegraphics[width=1.0\columnwidth]{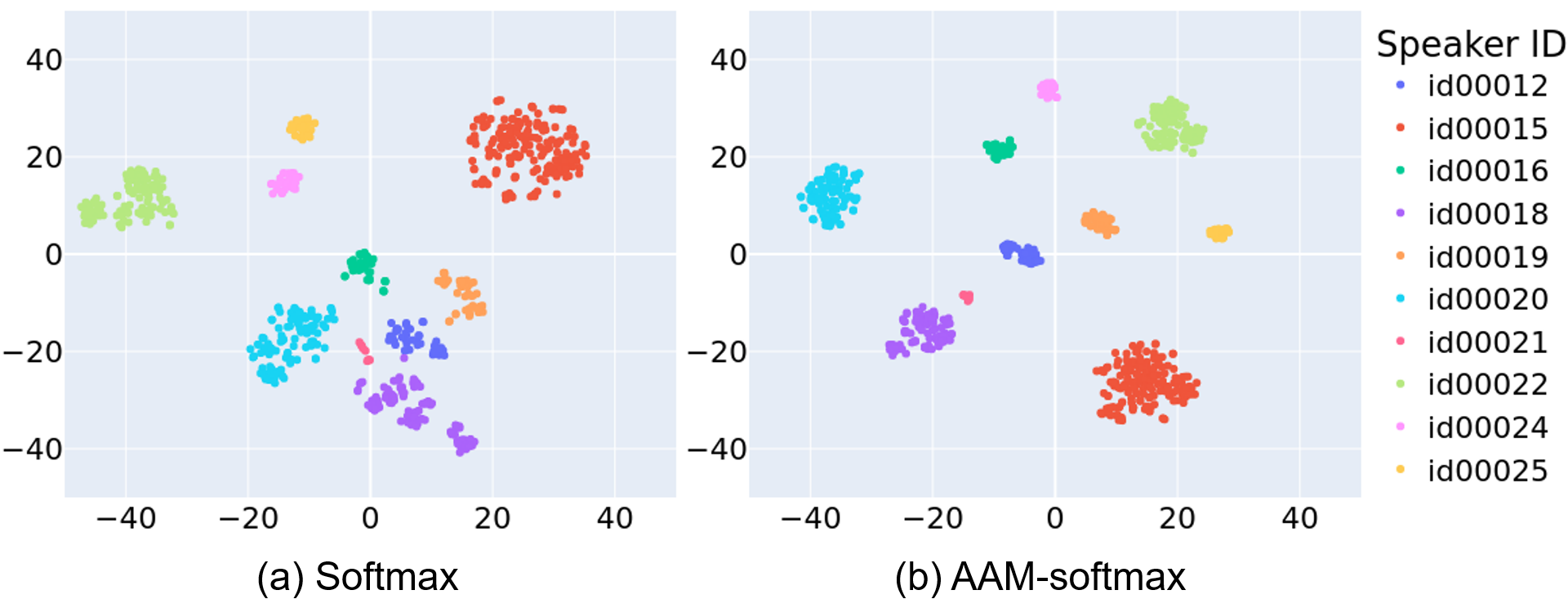}
\caption{\it t-SNE visualizations of (a)  conventional x-vector embeddings derived from a TDNN trained with softmax CE loss, and (b) large-margin embeddings derived with an ECAPA-TDNN trained with AAM-Softmax CE loss from the same 10 speakers. }
\label{fig:tsne}
\vspace*{-3mm}
\end{figure}
\section{Covariance Modeling for Large-Margin Embeddings}

PLDA \cite{prince07,ioffe06} was originally introduced in ASV to work with  i-vector framework \cite{dehak11,kenny10,matejka11}. Despite the i-vector front-end being replaced with more effective deep speaker embeddings, PLDA continues to be a promising back-end \cite{Villalba19,lee20}. 

We study empirically the between and within-speaker covariance of the conventional x-vector embeddings \cite{Snyder18} and large-margin embeddings from an ECAPA-TDNN\cite{Desplanques20}.
The plots in Fig.~\ref{fig:cov} (a) and (b) show that the within-speaker covariance of the conventional x-vector embeddings is larger than the between-speaker covariance in most of the dimensions, no matter whether length-normalization (LN) is applied. In contrary, the between-speaker covariance is larger than the within-speaker covariance for the large-margin embeddings in all the dimensions regardless of the LN application, as shown in Fig.~\ref{fig:cov} (c) and (d). It indicates that the use of large-margin softmax CE loss efficiently reduces the intra-speaker variability (enhanced intra-speaker compactness) in the embedding space. This motivates us to constrain PLDA models to match the reduced within-speaker variability in large-margin embeddings. In our implementation, we set the within-speaker covariance as a diagonal matrix in each iteration of the expectation-maximization (EM) \cite{em} steps in PLDA training. For the \emph{linear discriminant analysis} (LDA) pre-processing technique, we also use a constrained variant which keeps only the diagonal elements in the within-speaker covariance matrix calculated from the data in the calculation of the LDA transformation matrix. In this paper, they are referred to as LDA-diag and PLDA-diag.

Fig.~\ref{fig:tsne} show the t-SNE visualizations of the conventional and large-margin embeddings. Comparing the scatter plots in Fig.~\ref{fig:tsne} (a) and (b), it clearly shows the compactness of the individual classes with the large-margin embeddings with respect to the conventional x-vector embeddings. In addition, the between class distances are more uniform across classes with large-margin embeddings as shown in Fig.~\ref{fig:tsne}(b). This is consistent with Fig.~\ref{fig:cov} where the between-speaker covariance of the large-margin embeddings are distributed more evenly across all of the dimensions, while in the conventional embeddings, high covariance values concentrate in certain dimensions only.



\hskip -0.5cm
\section{Experiments}
\label{sec:exp}
\subsection{Experimental settings}
In order to verify the effectiveness of back-end techniques, 
the experiments are conducted on both VoxCeleb1 \cite{Nagrani17} and 
the Speaker in the Wild (SITW) core-core \cite{McLaren16} test sets.
For VoxCeleb1, we have exploited the original test set Vox1-o and the hard test set Vox1-h.
All of our front-ends and parametric back-ends are trained on VoxCeleb2 dataset \cite{Chung18}. 
Approximately $2\%$ of the train set is reserved for validation. 
Between our training and evaluation sets, there are no overlapping speakers.
%
We employ augmentation techniques to produce a variety of the training data for the embedding networks,
including random drops of audio chunks and frequency bands \cite{Park19}, speed perturbation \cite{Ko15}, 
environmental corruptions with a collection of room impulse responses (RIRs) and noise \cite{Ko17}.
For the parametric back-end training, a subset of VoxCeleb2 that consists of 300k utterances from 5,985 speakers  is used with no augmentation,  
considering the training and testing data are in similar conditions.

We study several systems of state-of-the-art TDNN, ECAPA-TDNN and MFA-TDNN backbones   
with softmax, AAM-Softmax and AM-Softmax cross-entropy (CE) losses for comparisons \cite{lee21,Liu22,Desplanques20,Snyder18}. The pooling options are average and attentive statistics pooling and posterior inference pooling \cite{lee21}.
The details of combinations are shown in Table~\ref{tab:exp1}.
We use SpeechBrain open-source toolkit\cite{sb21} to implement all the front-ends and extract speaker embeddings.
At the input of the neural networks, our systems utilize 80-dimensional filterbank features.

We evaluate three scoring methods: cosine similarity, PLDA and PLDA-diag,
and also the effect of length normalization (LN) \cite{Garcia-Romero11} and LDA as pre-processing steps for PLDA, as well as LDA-diag.
The  dimensions of LDA and LDA-diag are set to 150. 
Results are reported in terms of equal error rate (EER) and the minimum normalized detection cost function (MinDCF) at $P_{target}= 10^{-2}$ and $C_{FA} = C_{Miss} = 1$.

\subsection{Results and analysis}

\newcommand{\bb}[1]{\parbox[t]{2mm}{\multirow{4}{*}{\rotatebox[origin=l]{270}{#1}}}}
{\renewcommand{\arraystretch}{1.1}
\begin{table*}[ht]
\centering
\caption{EER and minDCF of the evaluations of three back-ends: cosine similarity, PLDA and PLDA-diag 
with five sets of large-margin embeddings (S1-S5), and two sets of the conventional softmax embeddings (S6, S7), 
on the three test sets: vox1-o, vox1-h and SITW core-core.
The large-margin softmax includes AM- and AAM-Softmax CE losses.
The backbones of the networks include: TDNN, ECAPA-TDNN and MFA-TDNN \cite{Liu22}.
The pooling options are average and attentive statistics pooling and posterior inference pooling\cite{lee21}.
}    
\scalebox{0.8}{
\begin{tabular}{p{1cm}|*{3}{p{0.06cm}p{0.06cm}p{0.1cm}|}p{0.07cm}p{0.1cm} |p{1.1cm}p{1.1cm}p{1.2cm}*{2}{|p{1.2cm}p{1.1cm}p{1.2cm}}}
    \hline
\multirow{5}{*}{ID}  
& \multicolumn{3}{c|}{Backbone} 
& \multicolumn{3}{c|}{Pooling} 
& \multicolumn{3}{c|}{Loss} 
& \multicolumn{2}{c|}{Dim} 
& \multicolumn{3}{c|}{Vox1-o} 
& \multicolumn{3}{c|}{Vox1-h}
& \multicolumn{3}{c}{SITW}\\
\cline{2-21}
&\bb{TDNN} &\bb{ECAPA} &\bb{MFA}
&\bb{Avg} &\bb{Att} &\bb{Posterior}
&\bb{Softmax} &\bb{AM} &\bb{AAM}
&\bb{192} &\bb{512} &&& &&& &&\\
& &&& &&& &&& && Cos & PLDA & PLDA-diag  & Cos & PLDA & PLDA-diag & Cos & PLDA & PLDA-diag \\
& &&& &&& &&& && &&& &&& &&\\
\hline
S1 \cite{Desplanques20} & & \checkmark &     & & \checkmark  &  & & & \checkmark   & \checkmark&   
&1.28/0.177 & 1.91/0.261 & \textbf{1.18/0.157}
& 2.47/\textbf{0.241} & 3.75/0.331 & \textbf{2.29}/0.243
& 1.83/0.167 & 2.35/0.240 & \textbf{1.42/0.160}
\\
S2 & &\checkmark &  & & \checkmark&  & & & \checkmark & &\checkmark   
&1.21/0.145 & 2.53/0.247 & \textbf{1.08/0.129} 
&2.47/\textbf{0.248} & 4.54/0.418 & \textbf{2.32}/0.249
&1.61/0.175 & 2.95/0.297 &  \textbf{1.48/0.169}
\\
S3 & & \checkmark &  & & \checkmark &  & &\checkmark &  & \checkmark&   
&1.22/0.127 & 1.75/0.193 & \textbf{1.16/0.125} 
&2.57/\textbf{0.251} & 3.74/0.348 & \textbf{2.44}/0.252
&1.92/0.184 & 2.49/0.242 & \textbf{1.56/0.174}  
\\
S4 \cite{lee21} & & \checkmark &  & &  &\checkmark  & & &\checkmark  &\checkmark&   
&1.25/0.150 & 1.93/0.208 & \textbf{1.16/0.136}
&2.43/\textbf{0.238} & 3.75/0.334 & \textbf{2.27}/0.239
&1.78/0.167 & 2.45/0.235 & \textbf{1.39/0.158}
\\
S5 \cite{Liu22}& & & \checkmark & &\checkmark &  & & & \checkmark &\checkmark&   
& 1.14/0.132 &1.56/0.208 & \textbf{1.02/0.114}
&2.26/0.225 &3.35/0.299 &\textbf{2.09/0.218} 
& 1.56/0.156 & 2.07/0.229 & \textbf{1.28/0.145} 
\\
S6 & & \checkmark&  & &\checkmark &  &\checkmark & &  &\checkmark&   
& 3.41/0.389  &\textbf{2.46/0.283}  & 2.76/0.298
&5.88/0.497   & \textbf{4.36/0.394} & 4.73/0.428 
&3.77/0.367   & 3.08/0.315 &  \textbf{2.71/0.313}
\\
S7 \cite{Snyder18} & \checkmark& &  & \checkmark& &  &\checkmark & &  &&   \checkmark
& 6.86/0.637  & \textbf{3.23/0.368} & 5.81/0.505
&12.42/0.778 & \textbf{5.87/0.505} &9.97/0.643 
&13.72/0.892 & \textbf{5.60/0.609} & 13.22/0.964
\\
\hline                    
   \end{tabular}}
    \label{tab:exp1}
\end{table*}}

We first investigate the intra-speaker compactness in the conventional softmax embeddings (S6-S7 in Table~\ref{tab:exp1}) and the
large-margin embeddings (S1-S5), respectively.
Only LN is used before scoring as the pre-processing step. 
As shown in Table~\ref{tab:exp1}, for both S6 and S7, PLDA outperforms cosine similarity measure,
while for the five systems (S1-S5) with different types of large margin softmax CE losses, cosine similarity measure achieves better performance than PLDA.
These observations are consistent on all three evaluation sets.
This indicates that the within-speaker variability in the conventional softmax embeddings are effectively reduced by channel compensation in PLDA, 
while the channel compensation is no longer essential for large-margin embeddings and even deteriorates the ASV performance.
Figure~\ref{fig:cov} depicts the difference in the covariance plots between different embeddings. 
Both the results in Tabel~\ref{tab:exp1} and the covariance plots show that the use of large-margin softmax CE loss efficiently reduces the intra-speaker variability in the embeddings.
Comparing the front-ends, the large-margin embeddings (S1-S5) achieve much better performance than the conventional embeddings (S6, S7), which also confirms the efficiency of large-margin softmax in learning speaker-discriminative embeddings.

\begin{figure}[t]
\centering
\includegraphics[width=1.2\columnwidth]{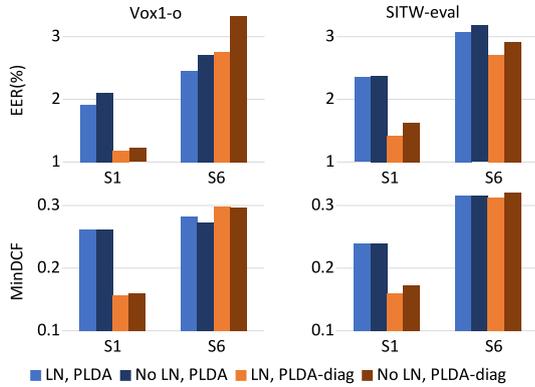}
\caption{{\it Comparisons of EER and minDCF when using length normalization or not in embedding pre-processing}}
\label{fig:ln}
\end{figure}
\begin{figure}[t]
\centering
\includegraphics[width=1.2\columnwidth]{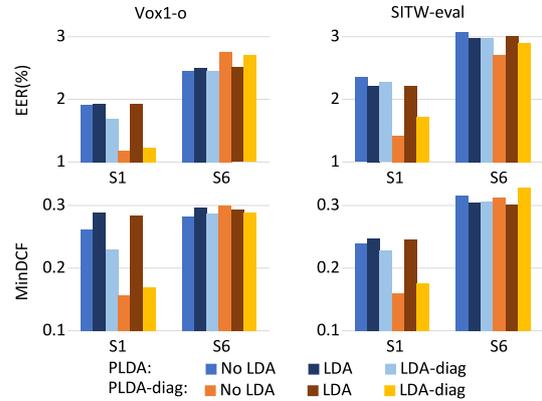}
\caption{{\it Comparisons of EER and minDCF when using LDA or not in embedding pre-processing}}
\label{fig:lda}
\end{figure}

Next, we investigate the effectiveness using diagonal within-class covariance matrix (denoted as PLDA-diag) in Table~\ref{tab:exp1}.
The use of the diagonalized within-speaker covariance in the PLDA model on the large-margin embeddings (S1-S5) reduces EER and minDCF on average by $40.8\%$ and $35.1\%$, respectively, compared with the conventional PLDA with full within-class covariance matrix. 
Additionally, it outperforms cosine similarity consistently, reducing EER and minDCF on average by $   10.9\%$ and $4.9\%$, respectively. 
For conventional embeddings (S6, S7),  on the contrary,
PLDA-diag degrades both EER and minDCF compared with the conventional PLDA.

Taking ECAPA-TDNN as a front-end example of large-margin embeddings (S1 vs. S2), we further investigate the effect of embedding dimensions on ASV performance. 
Cosine similarity gives similar performance across the two dimensional embeddings, 
while the degradation  produced by using the conventional PLDA in the 512-d embedding system S2 is almost double that of the 192-d embedding system S1.
If we use PLDA-diag instead of PLDA, the performance improves and both systems have similar performance again.

Next we investigate the feasibility of the pre-processing techniques of PLDA 
on the large-margin embeddings.
Since Vox1-h shows the same trend in the performance as Vox1-o, we exclude it considering the page limit.
Figure~\ref{fig:ln} shows the effect of length normalization (LN) on the large-margin embeddings (S1)  and the conventional embeddings (S6) with both PLDA and PLDA-diag back-ends on the Vox1-o and SITW core-core test sets.
We observe that applying LN reduces both EER and minDCF in almost all systems.
The performance improvement in EER is larger than that in minDCF. 
Therefore, we conclude that LN is still effective for large-margin speaker embeddings.
We also note that with or without LN, PLDA-diag outperforms PLDA significantly.
We have validated all the large-margin embeddings in Table~\ref{tab:exp1} and obtained the same results.

Figure~\ref{fig:lda} shows the effect of LDA pre-processing technique on the same front-ends and back-ends.
For the large-margin embeddings (S1), the use of the conventional LDA does not help in conventional PLDA systems, but drastically increases errors when applying to PLDA-diag systems.
Applying LDA-diag to the PLDA systems improves the performance, however, much less than the improvement brought by using PLDA-diag directly. 
Applying it to the PLDA-diag system degrades the performance slightly.
We conclude that for large-margin embeddings, 
removing the off-diagonal elements in the within speaker-covariance matrix in either LDA or PLDA improves speaker modeling. 
Using only PLDA-diag without LDA is sufficient to achieve good performance.
For the conventional embeddings (S6), applying both LDA and LDA-diag does not greatly affect the performance. LDA helps when there is a slight mismatch between the SITW test set and the model training set.

\section{Conclusions}

 This paper, for the first time, experimentally investigated the reasons of the shift from parametric back-ends to a simpler cosine similarity measure for the scoring of large-margin speaker embedding in speaker verification.
 Our experiments on the state-of-the-art ECAPA-TDNN networks with AAM-Softmax and AM-Softmax cross-entropy losses
 on VoxCeleb1 and SITW core-core test sets 
showed substantial increment in intra-speaker compactness making the conventional PLDA superfluous,
while the cosine similarity scoring seems to be sufficient. 
We found that simply discarding off-diagonal elements in the within-speaker covariance matrix of the PLDA model improved the performance significantly
with an average of $40.8\%$ EER reduction and $35.1\%$ minDCF reduction.
It also outperformed cosine scoring consistently with reductions in EER and minDCF by $10.9\%$ and $4.9\%$, respectively. 
In addition, this paper revisited the pre-processing techniques which have been widely used in the ASV back-ends in the past, and assessed their effects.
In the future, we will investigate the evaluations in mismatch domains.

\section{Acknowledgements}

This project is supported by the Agency of Science, Technology and Research (A$^\star$STAR), Singapore (Project No. CR-2021-005).

\bibliographystyle{IEEEtran}

\bibliography{mybib}


\end{document}